\documentclass[flushrt,preprint]{aastex}% onecolumn (second format,not aastex61)
\usepackage{graphicx}
\usepackage{url}
\usepackage{amsmath,amssymb}
\usepackage{natbib}
\begin{document}
\sloppy

\title{Changing World Extreme Temperature Statistics}
\shorttitle{World Temperature Extremes}
\shortauthors{Finkel \& Katz}
\author{J.~M.~Finkel}
\affil{Department of Physics, Washington University, St. Louis, Mo. 63130}
\author{J.~I.~Katz}
\affil{Department of Physics and McDonnell Center for the Space Sciences,\\
Washington University, St. Louis, Mo. 63130}
\affil{Tel.: 314-935-6276; Facs: 314-935-6219}
\email{katz@wuphys.wustl.edu}
\begin{abstract}
We use the Global Historical Climatology Network--daily database to
calculate a nonparametric statistic that describes the rate at which
all-time daily high and low temperature records have been set in nine
geographic regions (continents or major portions of continents) during
periods mostly from the mid-20th Century to the present.  This statistic was
defined in our earlier work on temperature records in the 48 contiguous
United States.  In contrast to this earlier work, we find that in every
region except North America all-time high records were set at a rate
significantly (at least $3\sigma$) higher than in the null hypothesis of a
stationary climate.  Except in Antarctica, all-time low records were set at
a rate significantly lower than in the null hypothesis.  In Europe, North
Africa and North Asia the rate of setting new all-time highs increased
suddenly in the 1990's, suggesting a change in regional climate regime; in
most other regions there was a steadier increase.
\end{abstract}
\keywords{climate change --- global warming --- extreme temperatures ---
extreme temperature indices}
%\maketitle

\section{Introduction}
The statistics of extreme events are not determined by mean trends.  The
fact that global, and most local, mean temperatures are increasing does not
by itself determine how frequently extreme events, such as all-time record
temperatures, occur.  Yet the extremes may, for practical purposes, be more
important than the mean trend
\citep{KB92,K93,N95,E97,E00,M00,F02,KT03,A06,P08,M09,AK10,RC11,L16}.
It has been predicted \citep{MT04,SH10,S16,IPE17} that an increasing rate of
temperature extremes will have significant effects on human life.  Model
predictions need to be supplemented by empirical data on the actual history
of extreme temperatures.  Although historical data cannot be directly
extrapolated to the future, they provide evidence of the past effects of
climate change on temperature extremes and can test the ability of the
models to predict (or ``retrodict'') climate beyond the mean warming trend.

In an earlier paper \citep{FK17} we defined an index that describes the
frequency of all-time record daily temperature highs and lows.  The value of
this nonparametric statistic tests the null hypothesis of no change in the
frequencies of record highs and lows and parametrizes rates of change of
their frequencies.  The index is based on the principle that in a stationary
time series with $N$ elements the probability that the last element sets an
all-time record is $1/N$.

We define a temperature extreme index as the sum over sites $i$ and calendar
days $j$ of $N_{i,j}$ if that year sets an all-time record for that site
and day and of zero if it doesn't, where $N_{i,j}$ is the number of years
in the data record for that site and calendar day.  The average (over all
sites and calendar days) of this index has a mean value of unity if there
are no systematic trends in the rates of daily extremes, more than unity if
more records are being set than for stationary statistics, and less than
unity if fewer records are being set.  This index has the advantages of
simplicity, an obvious mean value in the null hypothesis of stationary
climate and robustness against spurious extreme data.  Even an impossibly
high or low recorded value at one site and day has only a slight effect on
the index, unlike its effect on statistics like the variance that are
sensitive to outliers.  The index is applicable even to climatic databases
in which stations enter and leave the record irregularly and there are many
missing data.  It combines information from sites with temperature records
of varying lengths, and is not corrupted by missing data.

The index is computed for every site in the database and for every day of
the calendar for a given year.  We average the index over all sites and
calendar days for that year to minimize statistical variations.  The index
is then shown as a function of year, its mean (over years) found and a
linear trend fitted.  The index can be computed for any subset of sites and
calendar days, permitting study of regional or seasonal differences.  We
searched for seasonal effects, but found none that were significant,
although the larger statistical fluctuations mean that this is not a
sensitive test.

The reader should consult the earlier paper for details, a general
introduction and application to temperature records from the 48 contiguous
United States.  Here we extend that study to world-wide data from the Global
Historic Climate Network \citep{GHCN}.  This is important because temperature
extremes and their effects are predicted to be more severe outside North
America and because North America is only a small part of the world-wide
climate system.  The GHCN data have been the subject of careful quality
control and are described in detail by \cite{MDVGH12}.
\section{Results}
\begin{figure}
\centering
\includegraphics[width=0.9\textwidth]{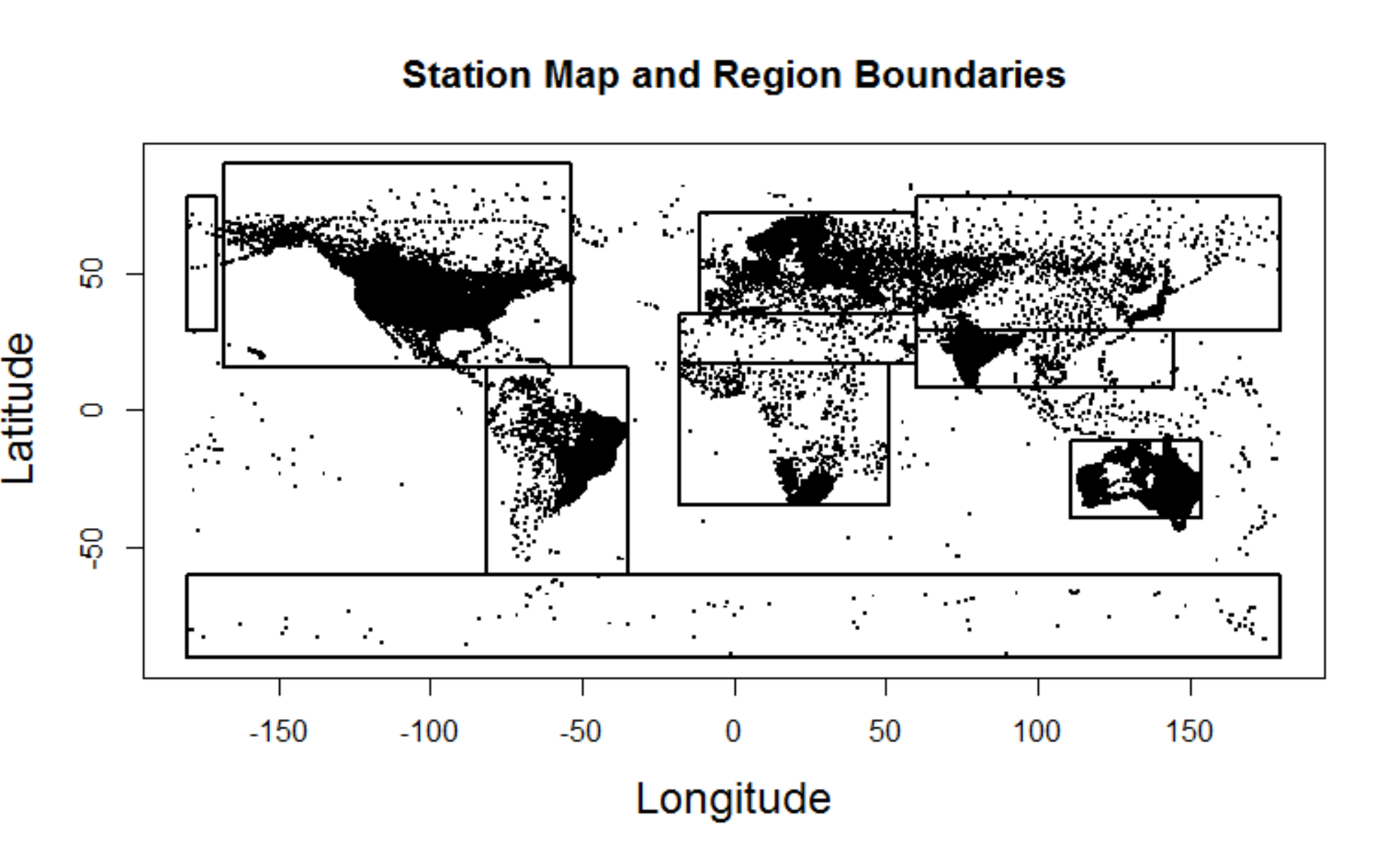}
\caption{\label{map} Map of all station locations and the nine regions.
Each dot represents a single station and the boxes enclose the regions
defined in Table~\ref{results}.  North Asia is divided between two boxes,
one on each side of 180 E/W.}
\end{figure}

We divide the world's land area, somewhat arbitrarily, into nine regions
(Fig.~\ref{map}).  Our results are shown in Table~\ref{results}.

\begin{table}
\centering
\begin{tabular}{|l|cc|cc|c|}
\hline
Name & Latitudes & Longitudes & High Index & Low Index & Epoch\\
\hline
Antarctica & 60--90 S & All & $1.071 \pm 0.022$ & $0.985 \pm 0.021$ & 1957--2016\\
Australia & 10.93--39.1 S & 111--154 E & $1.107 \pm 0.036$ & $0.896 \pm 0.020$ & 1957--2016\\
Europe & 35--72 N & 11.55 W--60 E & $1.170 \pm 0.038$ & $0.857 \pm 0.025$ & 1940--2016\\
N. America & 15.5--90 N & 53.5--168 W & $1.027 \pm 0.018$ & $0.918 \pm 0.016$ & 1893--2016\\
N. Asia & 29--77.7 N & 60 E--170.4 W & $1.147 \pm 0.033$ & $0.815 \pm 0.028$ & 1936--2016\\
N. Africa & 16.7--35 N & 17.7 W--60 E & $1.141 \pm 0.042$ & $0.820 \pm 0.029$ & 1945--2016\\
S. America & 15.5 N--60 S & 34.9--81.7 W & $1.175 \pm 0.039$ & $0.858 \pm 0.022$ & 1963--2016\\
S. Asia & 8.15--29 N & 60--145 E & $1.222 \pm 0.059$ & $0.848 \pm 0.026$ & 1951--2016\\
Sub-Sahara & 34.8 S--16.7 N & 17.7 W--51.2 E & $1.261 \pm 0.056 $ & $0.916 \pm 0.019$ & 1945--2016\\
\hline 
\end{tabular}
\caption{\label{results} Mean values of indices of all-time high and low
temperatures for the sites and epochs indicated.  Uncertainties are
$1\sigma$ errors of means and reflect the fact thta temperature variability
is strongly correlated from site to site and from day to day.  The numbers
of sites in the database rise rapidly at well-defined historical epochs that
differ for different regions.  The analyses of each region begin when the
database contains about 20\% of the maximum number of sites in that region
in order to avoid giving undue influence to early years with only a small
number of sites (and to those sites).  The (necessarily somewhat arbitrary)
dividing line between Europe and Asia is close to the Urals; North and South
Asia are separated at the latitude of Mt.~Everest and the Himalayas; North
Africa is separated from Sub-Saharan Africa in the Sahel.}
\end{table}

\begin{figure}
\centering
\includegraphics[width=0.95\textwidth]{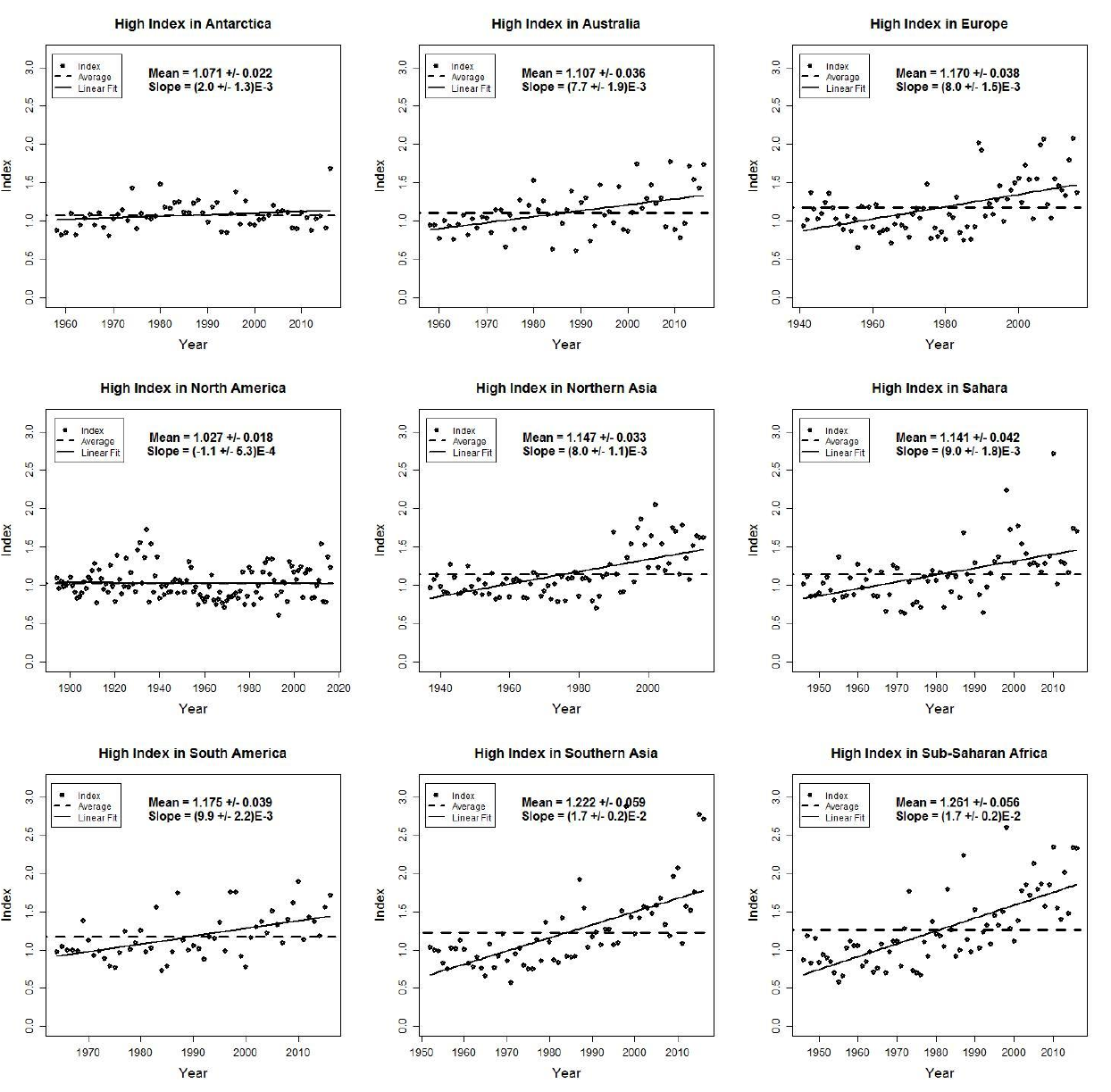}
\caption{\label{high} Histories of all-time high indices in each of nine
regions.}
\end{figure}

\begin{figure}
\centering
\includegraphics[width=0.95\textwidth]{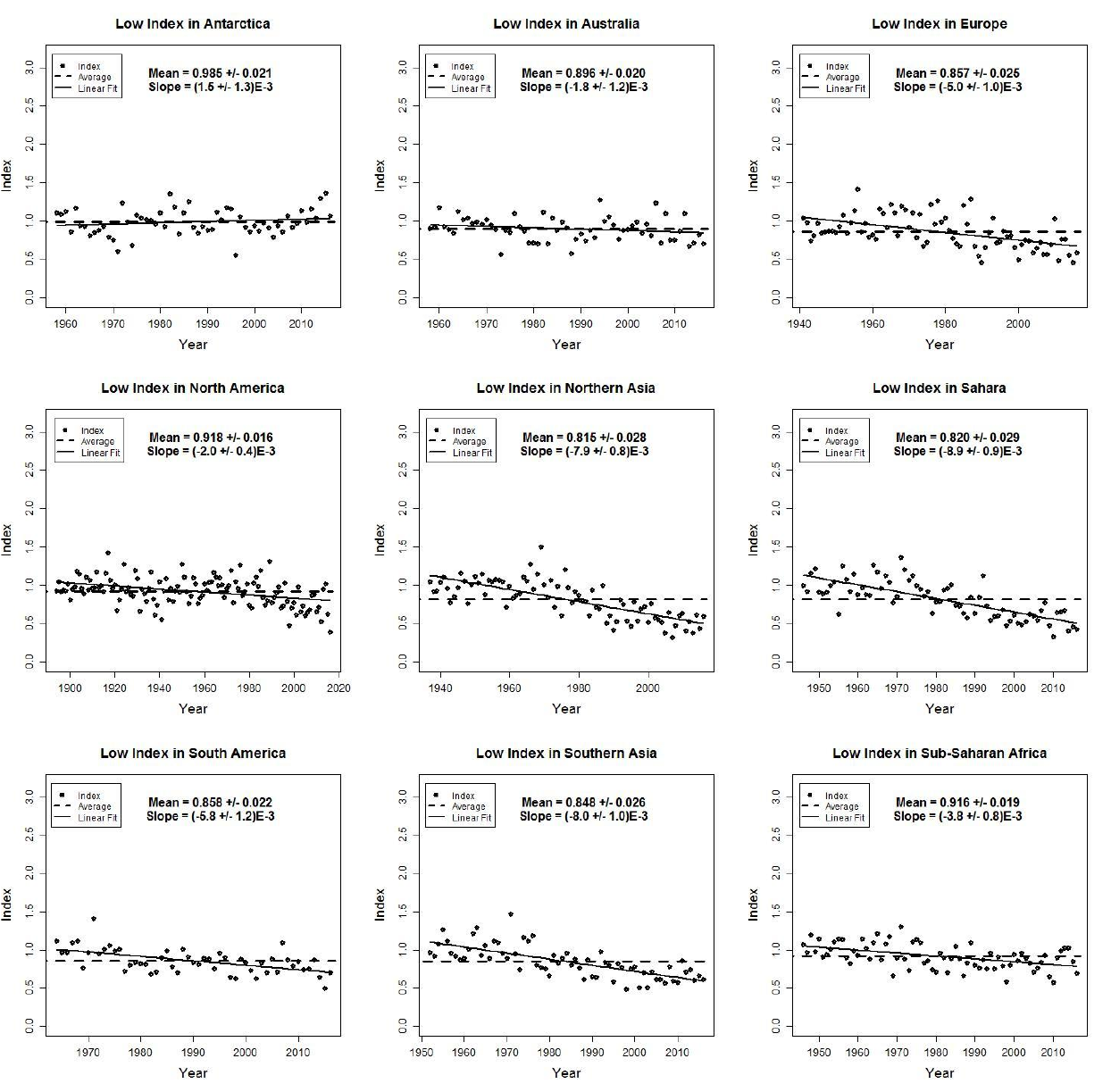}
\caption{\label{low} Histories of all-time low indices in each of nine
regions.}
\end{figure}

The time dependences of the high indices are shown in Fig.~\ref{high} and
those of the low indices are shown in Fig.~\ref{low}.
Eight of the nine regions show a mean high index that exceeds unity by
more than $3\sigma$, indicating a significantly increasing rate of record
highs over the period indicated.  The only exception is North America, for
which no significant trend is found (the nominal result is $1.5\sigma$).
This is consistent with our previous result \citep{FK17} for the 48
contiguous United States, that contribute most of the North American data,
of no significant change in this index.

Eight of the nine regions show a mean low index that is less than unity by
more than $3\sigma$, indicating a significantly decreasing rate of record
lows over the period indicated.  The North American result is consistent
with our previous result for the 48 contiguous United States.  The only
exception to the trend of decreasing record lows is Antarctica, for which
no significant trend is found.

\section{Discussion}
It is not surprising that a generally warming climate is accompanied by
an increasing frequency of all-time record daily high temperatures and a
decreasing frequency of all-time record daily low temperatures.  Perhaps
unexpected was the magnitude of the regional differences.  North America
is unique among the nine regions in showing no significant increase in the
rate of daily all-time highs.  The rate of decrease of all-time lows in
North America is also significantly less than the rates of decrease in most
of the other eight regions, with the exceptions of Antarctica (where no
significant decrease is observed at all) and Australia and sub-Saharan
Africa, where they decrease at approximately the same rate as in North
America.

Examination of the time-dependence of the indices in the Figures shows
additional information.  A slope in the mean index indicates a change in
the rate of climate change, although this cannot be quantified easily
because the index in any one year depends on values of the temperature in
the entire preceding record.  However, positive slopes of indices already
$> 1$, and negative slopes of indices already $< 1$, indicate speeding of
the rate of climate change.  This is evident for the high indices in Europe,
Northern and Southern Asia and sub-Saharan Africa, and to a lesser degree in
the low indices in several regions.  North America and Antarctica appear to
be exceptions to these trends.

Although there is much scatter in the annual values, the high indices in
Europe and Northern Asia may have undergone a transition to a regime of
systematically larger values and greater scatter in the 1990's.
\clearpage
\bibliography{finkelworld}

\begin{thebibliography}{}
\expandafter\ifx\csname natexlab\endcsname\relax\def\natexlab#1{#1}\fi

\bibitem[{Alexander {et~al.}(2006)Alexander, Zhang, Peterson, Caesar, Gleason,
  Tank, Haylock, Collins, Trewin, Rahimzadeh, Tagipour, Kumar, Revadekar,
  Griffiths, Vincent, Stephenson, Burn, Aguilar, Brunet, Taylor, New, Zhai,
  Rustucucci, \& Vazquez-Aguirre}]{A06}
Alexander, L.~V., Zhang, X., Peterson, T.~C., {et~al.} 2006, Global observed
  changes in daily climate extremes of temperature and precipitation, J.
  Geophys. Res., 111, D05109, doi:10.1029/2005JD006290

\bibitem[{Anderson \& Kostinski(2010)}]{AK10}
Anderson, A., \& Kostinski, A. 2010, Reversible record breaking and
  variability: temperature distributions across the globe, J. Appl. Meteor.
  Clim., 49, 1681

\bibitem[{Easterling {et~al.}(2000)Easterling, Meehl, Parmesan, Changnon, Karl,
  \& Mearns}]{E00}
Easterling, D.~R., Meehl, G.~A., Parmesan, C., {et~al.} 2000, Climate Extremes:
  Observations, Modeling and Impacts, Science, 289, 2068

\bibitem[{Easterling {et~al.}(1997)Easterling, Horton, Jones, Peterson, Karl,
  Parker, Salinger, Razuvayev, Plummer, Jamason, \& Folland}]{E97}
Easterling, D.~R., Horton, B., Jones, P.~D., {et~al.} 1997, Maximum and minimum
  temperature trends for the globe, Science, 277, 364,
  doi:10.1126/science.277.5324.364

\bibitem[{Finkel \& Katz(2017)}]{FK17}
Finkel, J.~M., \& Katz, J.~I. 2017, Changing U. S. Extreme Temperature
  Statistics, Int. J. Clim, in press

\bibitem[{Frich {et~al.}(2002)Frich, Alexander, Della-Marta, Gleason, Haylock,
  g.~Klein~Tank, \& Peterson}]{F02}
Frich, P., Alexander, L.~V., Della-Marta, P., {et~al.} 2002, Observed coherent
  changes in climatic extremes during the second half of the twentieth century,
  Clim. Res., 19, 193

\bibitem[{GHCN(2017)}]{GHCN}
GHCN. 2017, \url{ftp://ftp.ncdc.noaa.gov/pub/data/ghcn/daily}, accessed
  3/16/2017

\bibitem[{Im {et~al.}(2017)Im, Pal, \& Eltahire}]{IPE17}
Im, E.-S., Pal, J.~S., \& Eltahire, E. A.~B. 2017, Deadly heat waves projected
  in the densely populated agricultural regions of South Asia, Sci. Adv., 3,
  e1603322, dOI:10.1126/sciadv.1603322

\bibitem[{Karl {et~al.}(1993)Karl, Jones, Knight, Kukla, Plummer, Razuvayev,
  Gallo, Lindsay, Charlson, \& Peterson}]{K93}
Karl, T.~R., Jones, P.~D., Knight, R.~W., {et~al.} 1993, A new perspective of
  the recent global warming: asymmetric trends of daily maximum and minimum
  temperature, Bull. Am. Meteorol. Soc., 74, 1007

\bibitem[{Katz \& Brown(1992)}]{KB92}
Katz, R.~W., \& Brown, B.~G. 1992, Extreme events in a changing climate:
  variability is more important than averages, Clim. Chang., 21, 289,
  doi:10.1007/bf00139728

\bibitem[{Klein~Tank \& K{\"o}nnen(2003)}]{KT03}
Klein~Tank, A. M.~G., \& K{\"o}nnen, G.~P. 2003, Trends in indices of daily
  temperature and precipitation extremes in Europe, 1946--1999, J. Clim., 16,
  3665

\bibitem[{Lewis {et~al.}(2016)Lewis, King, \& Perkins-Kirkpatrick}]{L16}
Lewis, S., King, A., \& Perkins-Kirkpatrick, S. 2016, Defining a new normal for
  extremes in a warming world, Bull. Amer. Meteor. Soc., in press,
  doi:10.1175/BAMS-D-16-0183.1

\bibitem[{Meehl \& Tebaldi(2004)}]{MT04}
Meehl, G.~A., \& Tebaldi, C. 2004, More intense, more frequent, and
  longer-lasting heat waves in the 21st century, Science, 305, 994

\bibitem[{Meehl {et~al.}(2009)Meehl, Tebaldi, Walton, Easterling, \&
  McDaniel}]{M09}
Meehl, G.~A., Tebaldi, C., Walton, G., Easterling, D., \& McDaniel, L. 2009,
  Relative increase of record high maximum temperatures compared to record low
  minimum temperatures in the U.S., Geophys. Res. Lett., 36, L23701,
  doi:10.1029/2009GL040736

\bibitem[{Meehl {et~al.}(2000)Meehl, Zwiers, Evans, Knutson, Mearns, \&
  Whetton}]{M00}
Meehl, G.~A., Zwiers, F., Evans, J., {et~al.} 2000, Trends in Extreme Weather
  and Climate Events: Issues Related to Modeling Extremes in projections of
  Future Climate Change, Bull. Am. Met. Soc., 81, 427

\bibitem[{Menne {et~al.}(2012)Menne, Durre, Vose, Gleason, \&
  Houston}]{MDVGH12}
Menne, J.~M., Durre, I., Vose, R.~S., Gleason, B.~E., \& Houston, T.~G. 2012,
  An overview of the global historical climatology network---daily database, J.
  Atmos. Ocean. Tech., 29, 897, doi:10.1175/JTECH-D-11-00103.1

\bibitem[{Nicholls(1995)}]{N95}
Nicholls, N. 1995, Long-term climate monitoring and extreme events, Clim.
  Chang., 31, 231, doi:10.1007/bf01095148

\bibitem[{Platova(2008)}]{P08}
Platova, T.~V. 2008, Annual air temperature extrema in the Russian federation
  and their climatic changes, Russ. Meteorol. Hydrol., 33, 735

\bibitem[{Rahmstorf \& Coumou(2011)}]{RC11}
Rahmstorf, S., \& Coumou, D. 2011, Increase of extreme events in a warming
  world, Pub. Natl. Acad. Sci., 108, 17905

\bibitem[{Sch{\"a}r(2016)}]{S16}
Sch{\"a}r, C. 2016, Climate extremes: the worst heat waves to come, Nat. Clim.
  Change, 6, 128

\bibitem[{Sherwood \& Huber(2010)}]{SH10}
Sherwood, S.~C., \& Huber, M. 2010, An adaptibility limit to climate change due
  to heat stress, Proc. Natl. Acad. Sci. U. S. A., 107, 9552

\end{thebibliography}
\end{document}